\documentstyle[12pt,dina4,amsfont12]{article}

\input psfig.sty

\begin{document}
\newcommand{\bPsi}{{\bf \Psi} }
\newcommand{\bPhi}{{\bf \Phi} }
\newcommand{\dts}{\stackrel{\cdot}{*}}
\newcommand{\bU}{{\bf U} }
\newcommand{\bV}{{\bf V} }
\newcommand{\bz}{{\bf 0} }
\newcommand{\bA}{{\bf A} }
\newcommand{\bC}{{\bf C} }
\newcommand{\bmu}{{\mathcal U} }
\newcommand{\pl}[2]{\frac{\partial#1}{\partial#2}}
\newcommand{\Ta}{{\cal A}}
\newcommand{\Tb}{{\cal B}}
\newcommand{\Te}{{\cal E}}
\newcommand{\bu}{{\bf u} }
\newcommand{\p}{\partial}
\newcommand{\og}{\omega}
\newcommand{\Og}{\Omega}
\newcommand{\fl}[2]{\frac{#1}{#2}}
\newcommand{\dt}{\delta}
\newcommand{\tm}{\times}
\newcommand{\sm}{\setminus}
\newcommand{\nn}{\nonumber}
\newcommand{\ap}{\alpha}
\newcommand{\bt}{\beta}
\newcommand{\ld}{\lambda}
\newcommand{\Gm}{\Gamma}
\newcommand{\gm}{\gamma}
\newcommand{\vp}{\varphi}
\newcommand{\tht}{\theta}
\newcommand{\ift}{\infty}
\newcommand{\vep}{\varepsilon}
\newcommand{\ep}{\epsilon}
\newcommand{\kp}{\beta}
\newcommand{\Dt}{\Delta}
\newcommand{\Sg}{\Sigma}
\newcommand{\fa}{\forall}
\newcommand{\sg}{\sigma}
\newcommand{\ept}{\emptyset}
\newcommand{\btd}{\nabla}
\newcommand{\btu}{\Delta}
\newcommand{\tg}{\triangle}
\newcommand{\Th}{{\cal T}_h}
\newcommand{\ged}{\qquad \Box}
\newcommand{\bgv}{\Bigg\vert}
\renewcommand{\theequation}{\arabic{section}.\arabic{equation}}
\newcommand{\be}{\begin{equation}}
\newcommand{\ee}{\end{equation}}
\newcommand{\ba}{\begin{array}}
\newcommand{\ea}{\end{array}}
\newcommand{\bea}{\begin{eqnarray}}
\newcommand{\eea}{\end{eqnarray}}
\newcommand{\beas}{\begin{eqnarray*}}
\newcommand{\eeas}{\end{eqnarray*}}
\newcommand{\dpm}{\displaystyle}
\newtheorem{theorem}{Theorem}[section]
\newtheorem{lemma}{Lemma}[section]
\newtheorem{remark}{Remark}[section]

\newcommand{\Gmu}{\Gm_{_U}}
\newcommand{\Gml}{\Gm_{_L}}
\newcommand{\Gme}{\Gm_e}
\newcommand{\Gmi}{\Gm_i}
\newcommand{\lN}{{_N}}
\newcommand{\tld}[1]{\~ {#1}}
\newcommand{\td}[1]{\tilde{#1}}
\newcommand{\tp}{{\tilde{\phi}}}
\newcommand{\um}{\mu}
\newcommand{\bx}{{\bf x}}
\newcommand{\Eb}{E_{\beta,\Omega }}
\newcommand{\Ebb}{\epsilon_{\beta,\Omega }}
\newcommand{\mub}{\mu_{\kp_d} }
 \newcommand{\tu}{{\tilde u}}
 \newcommand{\Ez}{{E_0}}
 \newcommand{\Et}{{\tilde{E}}}
 \newcommand{\Vt}{{\tilde{V}_n(\bx)}}

\title{An efficient and spectrally accurate numerical method for
computing dynamics of rotating Bose-Einstein condensates}
\author{ {\it Weizhu Bao}\thanks{Email: bao@cz3.nus.edu.sg. Fax: 65-67746756,
URL: http://www.cz3.nus.edu.sg/\~{}bao/}
 \ \ and  {\it Hanquan
Wang}\thanks{Email: wanghanq@cz3.nus.edu.sg}\\
Department of Mathematics, National University of Singapore\\
Singapore 117543 }

\date{}
\maketitle

\begin{abstract}
In this paper, we propose an efficient and spectrally  accurate
numerical method for computing the dynamics of rotating
Bose-Einstein condensates (BEC) in two dimensions (2D) and 3D
based on the Gross-Pitaevskii equation (GPE) with an angular
momentum rotation term. By applying a time-splitting technique
for decoupling the nonlinearity and properly using the alternating
direction implicit (ADI) technique for the coupling in the angular
momentum rotation term in the GPE, at every time step, the GPE
in rotational frame is
decoupled into a nonlinear ordinary differential equation (ODE)
and two partial differential equations with constant coefficients.
This allows us to develop new time-splitting spectral (TSSP)
methods for computing the dynamics of BEC in a rotational frame.
The new numerical method is explicit, unconditionally stable,
and of spectral accuracy in
space and second order accuracy in time. Moreover, it is time
reversible and time transverse invariant, and conserves
the position density in the discretized level if the GPE does.
Extensive numerical
results are presented to confirm the above properties of the new
numerical method for rotating BEC in 2D \& 3D.

\end{abstract}


\section{Introduction}
\setcounter{equation}{0}

 Since the first experimental creation of a quantized vortex
in a gaseous Bose-Einstein condensate (BEC)
\cite{Madison,Abo-Shaeer,MAH,Raman}, there has been significantly
experimental and theoretical advances in the field of research
\cite{Adu,PB,Adhikari1,Bao2,Bao3,Bao5,Fetter,castin,Coddington,Dal,Fed2}.
Several experimental methods of vortex creation are currently in
use, including phase imprinting \cite{MAH,WH}, cooling of a
rotating normal gas \cite{Haljan}, and conversion of spin angular
momentum into orbital angular momentum by reversal of the magnetic
bias field in an Ioffe-Pritchard trap \cite{LGC,LSK,NIM}. The
topic of this paper is to propose an efficient and spectrally
accurate numerical method for studying quantized vortex dynamics
in a BEC by imposing a laser beam rotating with an angular
velocity on the magnetic trap holding the atoms to create a
harmonic anisotropic potential.

At temperatures $T$ much smaller than the critical condensation
temperature $T_c$, under mean field theory, the properties of a
BEC in a rotational frame are modelled by the well-known
time-dependent Gross-Pitaevskii equation (GPE) with an angular
momentum rotation term \cite{Adu,castin,Fed2}: \be\label{gpe1}
i\hbar\;\pl{\psi(\bx,t)}{t} =\left(-\fl{\hbar^2}{2m}\btd^2 +
V(\bx)+N U_0 |\psi|^2-\Omega L_z\right) \psi(\bx,t), \qquad
\bx\in{\Bbb R}^3, \ t\ge0, \ee where $\bx=(x,y,z)^T$ is the
Cartesian coordinate vector, $\psi(\bx,t)$ is the complex-valued
macroscopic wave function, $m$ is the atomic mass, $\hbar$ is the
Planck constant, $N$ is the number of atoms in the condensate,
$\Og$ is the angular velocity of rotating laser beam,
$V(\bx)=\frac{m}{2}\left(\omega_{x}^2 x^2+\omega_{y}^2 y^2
+\omega_{z}^2 z^2\right)$ with $\omega_{x}$, $\omega_{y}$ and
$\omega_{z}$  being the trap frequencies in $x$-, $y$- and
$z$-direction respectively. $U_0=\fl{4\pi \hbar^2 a_s}{m}$
describes the interaction between atoms in the condensate with
$a_s$ the $s$-wave scattering length,  and $L_z$ is the
$z$-component of the angular momentum. It is convenient to
normalize the wave function by requiring \be\label{norm}
\|\psi(\cdot,t)\|^2 :=\int_{{\Bbb R}^3} |\psi(\bx,t)|^2\; d\bx=1.
\ee

After proper nondimensionalization and dimension reduction, we can obtain
 the  following dimensionless GPE with an angular momentum rotation
term in the
 $d$-dimensions ($d=2,3$) \cite{Bao5,Bao3,Zhang}:
 \bea \label{gpeg}
&&i\;\pl{\psi(\bx,t)}{t}=-\fl{1}{2}\btd^2 \psi+ V_d(\bx)\psi +
\beta_d|\psi|^2\psi-\Omega L_z \psi, \qquad \bx\in {\Bbb R}^d,
\quad t\ge0, \qquad \\
\label{gpegp}
&&\psi(\bx,0)=\psi_0(\bx), \qquad \bx\in {\Bbb R}^d,
\qquad \hbox{with} \quad \|\psi_0\|^2:=\int_{{\Bbb R}^d}
|\psi_0(\bx)|^2\;d\bx=1,
\eea
 where $L_z=-i (x\p_y-y\p_x)$ and
 \be \label{uf}   V_d(\bx)=\left\{\ba{ll}
 \left(\gm_x^2x^2+\gm_y^2 y^2\right)/2, &\quad d=2, \\
 \dpm\left(\gm_x^2x^2+\gm_y^2 y^2+\gm_z^2 z^2\right)/2, &\quad d=3;\\
\ea\right. \ee with $\gm_x$, $\gm_y$ and $\gm_z$ being constants.

  In order to study effectively the dynamics of BEC, especially
in the strong repulsive interaction regime, i.e. $\bt_d\gg1$ in
(\ref{gpeg}), an efficient and accurate numerical method is one of
the key issues. For non-rotating BEC, i.e. $\Omega=0$ in
(\ref{gpeg}), many efficient and spectrally accurate numerical
methods were proposed in the literatures
\cite{BaoDP,BaoD,Bao2,Bao8,Bao9}, and they were
demonstrated that they are much better than the low-order finite
difference methods \cite{Cerimele,PS,MST,MFS}.
Thus  they were applied to  study
collapse and explosion of BEC in 3D \cite{Bao10} and
multi-component BEC \cite{Bao8} which are the very challenging
problems in numerical simulation of BEC. Due to the
appearance of the angular momentum rotation term in the GPE
(\ref{gpeg}), new numerical difficulties are introduced.
Currently, the numerical methods used in the physics literature
for studying dynamics of rotating BEC remain limited
\cite{Kasamatsu,Wang},
and they usually are low-order finite difference methods.
Recently, some efficient and accurate numerical methods were
designed for computing dynamics of rotating BEC. For example, Bao,
Du and Zhang \cite{Bao3} proposed a numerical method for computing
dynamics of rotating BEC by applying a time-splitting technique
for decoupling the nonlinearity in the GPE and adopting the polar
coordinates or cylindrical coordinates so as to make the
coefficient of the angular momentum rotation term constant. The
method is time reversible, unconditionally stable, implicit but
can be solved very efficiently, and conserves the total density.
It is of spectral accuracy in transverse direction, but usually of
second or
 fourth-order accuracy in radial direction.
Another numerical method is the leap-frog spectral method used for
studying vortex lattice dynamics in rotating BEC \cite{Zhang}. This
method is explicit, time reversible, of spectral accuracy in space
and second order accuracy in time. But it has a stability
constraint for time step \cite{Zhang}. The aim of this paper is to
develop a numerical method which enjoys advantages of both the
above two numerical methods. That is to say, the method is
explicit, unconditionally stable, time reversible, time transverse
invariant, and of spectral accuracy in space. We shall present
such an efficient, unconditionally stable and accurate numerical
method for discretizing the GPE in a rotational frame by applying
a time-splitting technique and an ADI technique, and constructing
appropriately spectral basis functions.

  The paper is organized as follows. In section 2,
we review some properties of GPE  in a rotational frame
(\ref{gpeg}) including conservation laws and analytical solutions
of condensate widths. In section 3, we present  a  new
time-splitting Fourier pseudospectral method  for
efficient and accurate simulation of GPE (\ref{gpeg})
in 2D \& 3D. In section 4, extensive numerical results are
reported to demonstrate the efficiency and spectral resolution in
space of our new numerical method.
Finally some conclusions are drawn in section 5.

\section{Some properties of the GPE}
 \label{1analytical}
 \setcounter{equation}{0}

For the convenience of the reader, in this section, we will review
some properties of the GPE with an angular momentum rotation term
(\ref{gpeg}), which will be used to test our new numerical method
proposed in the next section. First of all, the GPE (\ref{gpeg})
is time reversible and time transverse invariant. Second, it has
at least two important invariants which are  the {\sl
normalization of the wave function}
\begin{equation} \label{mass}
N(\psi)=\int_{{\Bbb R}^d}\; |\psi({\bf x}, t)|^2\; d{\bf x}\equiv
\int_{{\Bbb R}^d}\; |\psi({\bf x},0)|^2\; d{\bf x}=N(\psi_0)=1,
\qquad t\ge 0,
\end{equation}
and the {\sl energy}
\bea \label{energy}
E_{\bt,\Og}(\psi)&=&\int_{{\Bbb R}^d} \left[\fl{1}{2} \left|\btd
\psi\right|^2+ V_d(\bx)|\psi|^2 +\fl{\beta_d}{2}\; |\psi|^4-
\Omega \psi^* L_z \psi\right]d\bx\nonumber\\
&=&E_{\bt_d,\Og}(\psi_0), \qquad t\ge0, \eea
where $f^*$ and ${\rm
Im}(f)$ denote the conjugate and the imaginary part of the
function $f$ respectively. Third, it was proven that at least
for radial symmetric trap in 2D or cylindrical symmetric trap in
3D, i.e., $\gm_x=\gm_y$ in (\ref{gpeg}), the angular momentum
expectation and energy for non-rotating part are conserved
\cite{Bao3}, that is, for any given initial data $\psi_0(\bx)$ in
(\ref{gpegp}),
\be \label{consang} \langle L_z\rangle(t)\equiv
\langle L_z\rangle(0), \qquad E_{\bt,0}(\psi)
\equiv E_{\bt,0}(\psi_0), \quad t\ge0,
\ee
where the angular
momentum expectation which is a measure of the vortex flux is
defined as
\be \label{deflz} \langle
L_z\rangle(t):=\int_{\Bbb{R}^d} \psi^*(\bx,t) L_z \psi(\bx,t) \;
d\bx=i\int_{\Bbb{R}^d}\psi^*({\bx},t)(y\p_x-x\p_y)\psi({\bx},t)
d{\bx}, \quad t\geq 0.
\ee
Other very useful quantities
characterizing the dynamics of rotating BEC in 2D are the
condensate width defined as \be \label{defsigma} \dt_x(t) =
\int_{\Bbb{R}^d}x^2|\psi({\bx},t)|^2d{\bx}, \quad \dt_y(t) =
\int_{\Bbb{R}^d}y^2|\psi({\bx},t)|^2d{\bx}, \quad
\dt_r(t)=\dt_x(t)+\dt_y(t). \ee As proven in \cite{Bao3}, in 2D
with a radial symmetric trap, i.e., $d = 2$ and
$\gm_x=\gm_y:=\gm_r$ in (\ref{gpeg}), for any initial data
$\psi_0(x,y)$ in (\ref{gpegp}), we have for any $t\ge0$
\be
\label{solutiondtr} \dt_r(t) = \fl{E_{\bt,\Og}(\psi_0)+\Og\langle
L_z\rangle(0)}{\gm_r^2}\left[1-\cos(2\gm_rt)\right]
+\dt_r^{(0)}\cos(2\gm_rt)
 +\fl{\dt_r^{(1)}}{2\gm_r}\sin(2\gm_r t),
\ee
where $\dt_r^{(0)}: = \dt_r(0)=\dt_x(0)+\dt_y(0)$ and
$\dt_r^{(1)}:=\dot{\dt}_r(0)=\dot{\dt}_x(0)+\dot{\dt}_y(0)$ with
for $\ap=x$ or $y$
\beas &&\dt_{\ap}(0) = \dt_{\ap}^{(0)} =
\int_{\Bbb{R}^2}\ap^2|\psi_0({\bx})|^2
d{\bx}, \\
&&\dot{\dt}_{\ap}(0) = \dt_{\ap}^{(1)} = 2\int_{\Bbb{R}^2}\ap
\left[ -\Og |\psi_0|^2\left(x\p_y-y\p_x\right)\ap+ {\rm
Im}\left(\psi^*_0\p_\ap\psi_0\right)\right]\; d{\bx}.
 \eeas
Furthermore, when the initial data $\psi_0(x,y)$ in (\ref{gpegp})
satisfies \be \label{vortex_initial}
\psi_0(x,y)=f(r)e^{im\theta}\quad{\rm with}\quad m\in{\Bbb
Z}\quad{\rm and} \quad f(0) = 0\quad{\rm when}\quad m\neq0, \ee
with $(r,\tht)$ being the polar coordinates in 2D, we have, for
any $t\geq 0$, \bea \label{solution dt_xy}
\dt_x(t)&=&\dt_y(t) = \fl{1}{2}\dt_r(t)\nonumber\\
&=&\fl{E_{\bt,\Og}(\psi_0)+m\Og}{2\gm_x^2}\left[1-\cos(2\gm_xt)\right]
+\dt_x^{(0)}\cos(2\gm_xt)+\fl{\dt_x^{(1)}}{2\gm_x}\sin(2\gm_xt).
\qquad  \eea These immediately imply that $\dt_r(t)$ is a periodic
function with angular frequency doubling the trapping frequency in
2D with a radial symmetric trap, and also $\dt_x(t)$ and
$\dt_y(t)$ are periodic functions with frequency doubling the
trapping frequency provided that the initial data satisfies
(\ref{vortex_initial}).

\section{A time-splitting pseudospectral method for rotating BEC}
\setcounter{equation}{0}

In this section, we will present an explicit, unconditionally
stable and spectrally accurate numerical method to solve the GPE
(\ref{gpeg}) for dynamics of rotating BEC.

Due to the external trapping potential $V_d(\bx)$, the
   solution $\psi(\bx,t)$ of (\ref{gpeg})-(\ref{gpegp})
   decays to zero exponentially fast when $|\bx|\to \infty$.
Thus in practical computation, we always truncate the problem
(\ref{gpeg})-(\ref{gpegp}) into a bounded computational domain
with homogeneous Dirichlet boundary condition:
 \bea \label{GPE2}
&&i\;\p_t\psi({\bx},t) = -\fl{1}{2}\nabla^2\psi +
\left[V(\bx)-iW(\bx)\right] \psi +
\bt_d|\psi|^2\psi - \Og L_z\psi,\quad {\bx}\in \Og_{\bx},\
t>0,\qquad \\
\label{initial_data1} &&\psi(\bx,t) =0, \qquad \bx \in
\Gm=\p\Og_{\bx}, \qquad t\ge0, \\
\label{initial_data2}&&\psi({\bx},0) = \psi_0({\bx}), \quad
{\bx}\in \bar{\Og}_\bx;
\eea
where  $W(\bx)\ge0$ corresponds to a
localized loss term \cite{Simula} and $V(\bx)=V_d(\bx) +V_p(\bx)$
with $V_p(\bx)\ge0$ a conservative repulsive pinning potential
\cite{Simula}.
Here  we choose
$\Og_{\bx}=[a,b]\tm[c,d]$ in 2D, and resp.,
$\Og_{\bx}=[a,b]\tm[c,d]\tm[e,f]$ in 3D,
 with $|a|$, $b$, $|c|$, $d$, $|e|$  and $f$
sufficiently large. The use of more sophisticated radiation
boundary conditions is an interesting topic that remains to be
examined in the future.

\subsection{Time-splitting}

We choose a time step size $\btu t>0$. For $n=0,1,2,\cdots$,  from
time $t=t_n=n\btu t$ to $t=t_{n+1}=t_n+\btu t$, the GPE
(\ref{GPE2}) is first solved in two splitting steps. One solves
first \be \label{fstep} i\;\p_t\psi({\bx},t) = -\fl{1}{2}\nabla^2
\psi(\bx,t) - \Og L_z \psi(\bx,t)\ee for the time step of length
$\btu t$, followed by solving
 \be \label{sstep}
i\;\p_t\psi({\bx},t) = \left[V(\bx)-iW(\bx)\right] \psi(\bx,t) +
\bt_d|\psi(\bx,t)|^2\psi(\bx,t),
 \ee
for the same time step.
 For $t\in[t_n,t_{n+1}]$, multiplying (\ref{sstep}) by
$\psi^*$, the conjugate of $\psi$, we get
 \be \label{sstepi1}
i\;\psi^*(\bx,t)\p_t\psi({\bx},t) =
\left[V(\bx)-iW(\bx)\right] |\psi(\bx,t)|^2 +
\bt_d|\psi(\bx,t)|^4.
 \ee
Subtracting the conjugate of (\ref{sstepi1}) from (\ref{sstepi1})
and multiplying by $-i$, one obtains
 \be \label{sstepi2}
\frac{d}{dt}|\psi(\bx,t)|^2 = \psi^*\p_t \psi +\psi \p_t \psi^*
=-2W(\bx)|\psi(\bx,t)|^2.
\ee
Solving (\ref{sstepi2}), we get
\be\label{solut5}
|\psi(\bx,t)|^2 =e^{-2W(\bx)(t-t_n)}|\psi(\bx,t_n)|^2, \qquad
t_n\le t\le t_{n+1}.
\ee
Substituting (\ref{solut5}) into (\ref{sstep}), we obtain
 \be \label{sstepi3}
i\;\p_t\psi({\bx},t) = \left[V(\bx)-iW(\bx)\right] \psi(\bx,t) +
\bt_d e^{-2W(\bx)(t-t_n)} |\psi(\bx,t_n)|^2\psi(\bx,t).
 \ee
Integrate (\ref{sstepi3}) from $t_n$ to $t$, we
find for $\bx \in \Og_\bx$ and
$t_n\le t\le t_{n+1}$:
\be \label{solode}
\psi(\bx,t)=\left\{\ba{ll}
e^{-i[V(\bx)+\bt_d|\psi(\bx,t_n)|^2](t-t_n)}\;
\psi(\bx,t_n), &W(\bx)=0,\\
 \ \\
\frac{\psi(\bx,t_n)}{e^{W(\bx)(t-t_n)}}
e^{-i[V(\bx)(t-t_n)+\beta_d|\psi(\bx,t_n)|^2(1-e^{-2W(\bx)(t-t_n)})/2W(\bx)]},
&W(\bx)>0.
\end{array} \right.
\ee
To discretize (\ref{fstep}) in space, compared with
non-rotating BEC \cite{BaoDP,Bao2,BaoD}, i.e. $\Og=0$ in (\ref{gpeg}), the
main difficulty is that the coefficients in $L_z$ are {\sl not}
constants which cause big trouble in applying sine or Fourier
pseudospectral discretization. Due to the special structure in the
angular momentum rotation term $L_z$ (\ref{uf}), we will apply
the alternating direction implicit (ADI) technique and decouple
the operator $-\fl{1}{2}\nabla^2 - \Og L_z$ into two one dimensional
operators such that each operator becomes a summation of terms
with constant coefficients in that dimension. Therefore, they can
be discretized in space by Fourier pseudospectral method and
the ODEs in phase space can be integrated  analytically. The details for
discretizing (\ref{fstep}) in 2D \& 3D will be presented in the
next two subsections respectively.

\subsection{Discretization in 2D}

When $d=2$ in (\ref{fstep}), we choose mesh sizes $\btu x>0$ and
$\btu y>0$ with $\btu x=(b-a)/M$ and $\btu y=(d-c)/N$ for $M$ and
$N$ even positive integers, and let the grid points be
\[x_j=a + j \btu x, \quad j=0,1,2,\cdots, M; \quad y_k=c +k
\btu y, \quad k=0,1,2,\cdots,N.\] Let $\psi_{jk}^n$ be the
approximation of $\psi(x_j,y_k,t_n)$ and $\psi^n$ be the solution
vector with component $\psi_{jk}^n$.

  From time $t=t_n$ to $t=t_{n+1}$, we solve (\ref{fstep}) first
\be\label{fstep1} i\;\p_t\psi({\bx},t) = -\fl{1}{2}\p_{xx}
\psi(\bx,t) - i\Og y\p_x \psi(\bx,t), \ee for the time step of
length $\btu t$, followed by solving \be\label{fstep2}
i\;\p_t\psi({\bx},t) = -\fl{1}{2}\p_{yy} \psi(\bx,t) + i\Og x\p_y
\psi(\bx,t), \ee for the same time step. The detailed
discretizations of (\ref{fstep1}) and (\ref{fstep2})
 are shown in Appendix A.

\subsection{Discretization in 3D}

When $d=3$ in (\ref{fstep}), we choose mesh sizes $\btu x>0$, $\btu
y>0$ and $\btu z>0$ with $\btu x=(b-a)/M$, $\btu y =(d-c)/N$ and
$\btu z=(f-e)/L$ for $M$, $N$  and $L$ even positive integers, and
let the grid points be
\[x_j=a + j \btu x, \ 0\le j\le M; \quad y_k=c +k
\btu y, \ 0\le k\le N; \quad z_l=e+l \btu z, \ 0\le l\le L.\] Let
$\psi_{jkl}^n$ be the approximation of $\psi(x_j,y_k,z_l,t_n)$ and
$\psi^n$ be the solution vector with component $\psi_{jkl}^n$.

  Similar as those for 2D case, from time $t=t_n$ to $t=t_{n+1}$,
  we solve (\ref{fstep}) first
\be\label{fstep3} i\;\p_t\psi({\bx},t) =
\left(-\fl{1}{2}\p_{xx}-\frac{1}{4}\p_{zz} - i\Og y\p_x\right)
\psi(\bx,t), \ee for the time step of length $\btu t$, followed by
solving \be\label{fstep4} i\;\p_t\psi({\bx},t) =
\left(-\fl{1}{2}\p_{yy} -\frac{1}{4}\p_{zz} + i\Og x\p_y\right)
\psi(\bx,t), \ee for the same time step.
The detailed
discretizations of (\ref{fstep3}) and (\ref{fstep4}) are
 shown in Appendix B.

\subsection{Stability}

We define the usual discrete $l^2$-norm for the solution $\psi^n$
as \be\label{norm2d} \|\psi^n\|_{l^2}=\sqrt{\fl{b-a}{M}\
\frac{d-c}{N}\;\sum_{j=0}^{M-1}\;\sum_{k=0}^{N-1}\
\left|\psi_{jk}^n\right|^2}, \ee for $d=2$, and for $d=3$
\be\label{norm3d} \|\psi^n\|_{l^2}=\sqrt{\fl{b-a}{M}\
\frac{d-c}{N}\ \frac{f-e}{L}\; \sum_{j=0}^{M-1}\;
\sum_{k=0}^{N-1}\; \sum_{l=0}^{L-1}\ \left|\psi_{jkl}^n\right|^2}.
\ee

For the {\it stability} of the time-splitting spectral
approximations (\ref{tssp2d}) for 2D and (\ref{tssp3d}) for 3D, we
have the following lemma, which shows that the total density is
conserved when $W(\bx)\equiv 0$, and resp. decreased when
$W(\bx)>0$, in the discretized level.

\begin{lemma}\label{stability}
  The time-splitting spectral schemes (\ref{tssp2d}) for 2D and
(\ref{tssp3d}) for 3D GPE with an angular momentum rotation term
are unconditionally stable. In fact, for any mesh sizes $\btu
x>0$, $\btu y>0$ and $\btu z>0$, and time step size $\btu t>0$,
\be \label{stabu} \|\psi^{n}\|_{l^2}\le \|\psi^{n-1}\|_{l^2}
\le \|\psi^0\|_{l^2}=\|\psi_0\|_{l^2},
 \qquad n=1,2,\cdots\;.
\ee
In addition, if $W(\bx)\equiv 0$ in (\ref{GPE2}), then we have
\be \label{stabu1} \|\psi^{n}\|_{l^2}\equiv
\|\psi^0\|_{l^2}=\|\psi_0\|_{l^2},
 \qquad n=1,2,\cdots\;.
\ee
\end{lemma}

\noindent {Proof:}\ Follows the line of the analogous results for
the linear and nonlinear Schr\"{o}dinger equations in
\cite{BaoD,WSP1,WSP2,BaoDP,Bao9}.

\section{Numerical results}
\setcounter{equation}{0}

In this section, we first test the accuracy of our new numerical
method (\ref{tssp2d}) for 2D and (\ref{tssp3d}) for  3D  and
compare our numerical results with the analytical results reviewed
in section \ref{1analytical}. Then we apply our new numerical
method to study vortex lattice dynamics in rotating BEC by
changing the trapping frequencies and to generate a giant vortex
by introducing a localized loss term. Our aim is not to find
new physics phenomena but to demonstrate the efficiency and high resolution
of our new numerical method

\subsection{Accuracy test}

 To test the accuracy of our numerical method in 2D, we take
 $\beta_2=100$ and  $\Og=0.7$ in (\ref{gpeg}).
 The initial condition in (\ref{gpegp}) is taken as
 \be\label{init}
  \psi_0(x,y)=\fl{(\gm_x\gm_y)^{\fl{1}{4}}}{\pi^{\fl{1}{2}}}\ e^
 {-(\gm_x x^2+\gm_y y^2)/2}, \qquad \bx=(x,y)^T\in {\Bbb R}^2.
 \ee
 We take $\gm_x=1.0$ and $\gm_y=2.0$ in (\ref{gpeg}) and
 (\ref{init}). Similar example was used in \cite{Bao3,Zhang}
for testing numerical
accuracy of different numerical methods for rotating BEC.
 The GPE (\ref{gpeg}) is solved on $[-8,8]\times[-8,8]$,
 i.e. we take $a=-8$, $b=8$, $c=-8$ and $d=8$.
Let $\psi$ be the {\sl exact} solution which is obtained
numerically by using our method with a very fine mesh and small time
step, e.g. $\btu x = \btu y=\fl{1}{64}$ and $\btu t =0.0001$, and
$\psi^{(\btu x, \btu y, \btu t)}$ be the numerical solution
obtained with the mesh size $(\btu x, \btu y)$ and  time step
$\btu t$.

 First we test the spectral accuracy in space by choosing a very
 small time step $\btu t=0.0001$, and solving the problem for each
 fixed $\bt_2$ with different mesh size $\btu x=\btu y$ so that
 the  discretization errors in time can be neglected comparing to
 those in space. The errors $\|\psi(t) -\psi^{(\btu x,
\btu y, \btu t)}(t)\|_{l^2}$
 at $t=0.5$ are shown in Table 1 for different values $\bt_2$ and
 $h=\btu x=\btu y$.

\begin{table}[htb] \label{tab1}
 \begin{center}
\begin{tabular}{cccc}
  \hline
 h &1/2 &1/4 &1/8 \\  \hline
$\bt_2=20$ & 2.68E-2 & 6.5E-5 & 3.41E-10  \\
$\bt_2=50$ & 0.1315 & 2.01E-3 & 5.91E-8  \\
$\bt_2=100$ & 0.4287 & 1.94E-2 & 8.89E-6  \\
  \hline
\end{tabular}
\end{center}
Table 1: Spatial discretization errors $\|\psi(t)-\psi^{(\btu x,\btu y,\btu
t)}(t)\|_{l^2}$ at $t=0.5$ in  2D.
\end{table}

Next we test the second-order accuracy in time. Table 2 lists the
errors at $t=0.5$ for different values of $\bt_2$ and time steps
$\btu t$ with a fine mesh in space, i.e. $\btu x=\btu y=1/16$.

 \begin{table}[htb]\label{tab2}
 \begin{center}
 \begin{tabular}{ccccccc}
   \hline
  $\Delta t$    &1/40  &1/80  &1/160 &1/320 &1/640 \\   \hline
$\bt_2=20$ &7.86E-4 & 1.95E-4 & 4.86E-5 &1.21E-5 & 3.02E-6\\
$\bt_2=50$ &2.95E-3 & 7.24E-4 & 1.80E-4 & 4.49E-5 & 1.12E-5 \\
$\bt_2=100$   &7.98E-3 &1.98E-3 &4.76E-4 &1.19E-4 &2.96E-5\\
\hline
 \end{tabular}
\end{center}

Table 2: Temporal discretization errors $\|\psi(t)-\psi^{(\btu x,\btu y,\btu
t)}(t)\|_{l^2}$ at $t=0.5$ in 2D.
\end{table}

Similarly,  to test the accuracy of our numerical method in 3D, we
take
 $\beta_3=100$ and  $\Og=0.7$ in (\ref{gpeg}).
 The initial condition in (\ref{gpegp}) is taken as
 \be\label{init3d}
  \psi_0(x,y,z)=\fl{(\gm_x\gm_y\gm_z)^{\fl{1}{4}}}{\pi^{\fl{3}{4}}}\ e^
 {-(\gm_x x^2+\gm_y y^2+\gm_y z^2 )/2}, \qquad \bx=(x,y,z)^T\in {\Bbb R}^3.
 \ee
 We take $\gm_x=\gm_y=\gm_z=1.0$ in (\ref{gpeg}) and
 (\ref{init3d}), and solve
 the GPE (\ref{gpeg}) in 3D on $[-8,8]\times[-8,8]\times[-8,8]$.
Again let $\psi$ be the {\sl exact} solution which is obtained
numerically by using our method with a fine mesh and small time step,
e.g. $\btu x = \btu y=\btu z=\fl{1}{8}$ and $\btu t =0.0001$, and
$\psi^{(\btu x, \btu y, \btu z, \btu t)}$ be the numerical
solution obtained with mesh size $(\btu x, \btu y, \btu z)$
and time step $\btu t$.
Table 3 shows the spatial discretization errors
$\|\psi(t) -\psi^{(\btu x,
\btu y, \btu z, \btu t)}(t)\|_{l^2}$  at $t=0.5$ with $\btu t=0.0001$
 for different values $\bt_3$ and mesh sizes
 $h=\btu x=\btu y =\btu z$. Table 4 lists the
errors at $t=0.5$ for different values of $\bt_3$ and time steps
$\btu t$ with a fine mesh in space, i.e. $\btu x=\btu y=\btu
z=1/8$.

\begin{table}[htb] \label{tab3}
 \begin{center}
\begin{tabular}{cccc}  \hline
 h &1   &1/2 &1/4 \\
  \hline
  $\bt_3=20$ &5.78E-2 & 1.27E-3 & 2.38E-8  \\
  \hline
  $\bt_3=50$ &0.1515 & 9.50E-3 & 2.45E-6  \\
  \hline
$\bt_3=100$ &0.3075 & 3.88E-2 & 7.08E-5   \\
  \hline
\end{tabular}
\end{center}
Table 3: Spatial discretization errors $\|\psi(t)-\psi^{(\btu x,\btu y,\btu
,\btu z, t)}(t)\|_{l^2}$ at $t=0.5$ in  3D.
\end{table}

\begin{table}[htb]\label{tab4}
 \begin{center}
 \begin{tabular}{ccccccc}
   \hline
  $\Delta t$    &1/40  &1/80  &1/160 &1/320 &1/640 \\   \hline
  $\bt_3=20$ & 1.778E-4 & 4.435E-5 & 1.108E-5 & 2.767E-6 & 6.897E-7 \\
  \hline
  $\bt_3=50$ & 6.266E-4 & 1.559E-4 & 3.892E-5 & 9.718E-6 & 2.422E-6 \\
  \hline
  $\bt_3=100$   &1.63E-3 &4.0379E-4 &1.0069E-4 &2.5141E-5 &6.265E-6\\
\hline
 \end{tabular}
\end{center}

Table 4: Temporal discretization errors $\|\psi(t)-\psi^{(\btu x,\btu y,
\btu z,\btu t)}(t)\|_{l^2}$ at $t=0.5$ in  3D.
\end{table}

From Tables 1-4,  we can draw the following conclusions:
(i) The method (\ref{tssp2d}) or (\ref{tssp3d}) is of spectral
accuracy in space and second order accuracy in time. (ii) For a
given fixed mesh size and time step, when $\bt_d$ is increasing,
the errors are increasing too. This implies that when the number
of atoms in the condensate is increasing, i.e. $\bt_d$ is
increasing, more grid points and smaller time step are needed in
practical computation in order to achieve a given accuracy.

  Furthermore, Fig. 1 shows time evolution of
  the normalization $N(\psi)(t)$,
 energy $E_{\bt,\Og}(\psi)(t)$, angular momentum expectation
 $\langle L_z\rangle(t)$ and condensate widths for the above parameters setup
 in 2D and 3D.

\begin{figure}[t!]  \label{fig1}
\centerline{(a)\psfig{figure=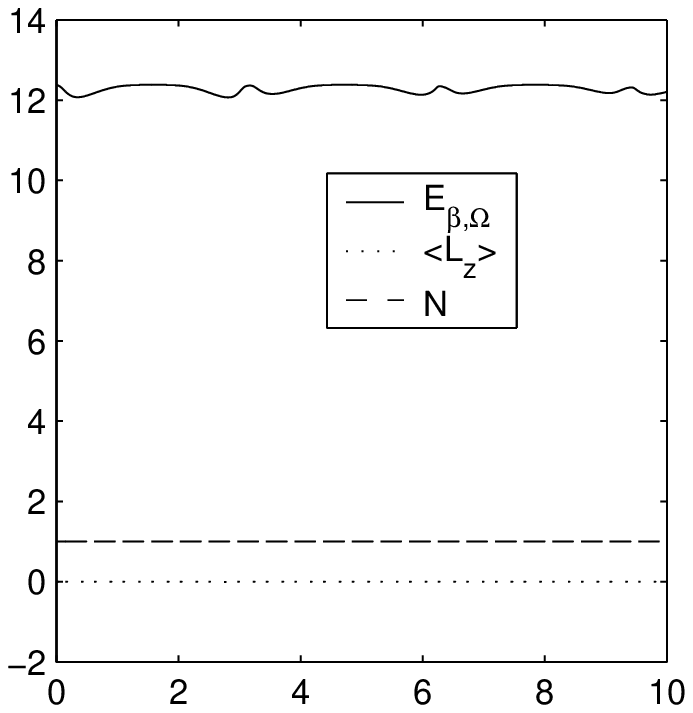,height=4cm,width=6cm,angle=0}
\qquad (b) \psfig{figure=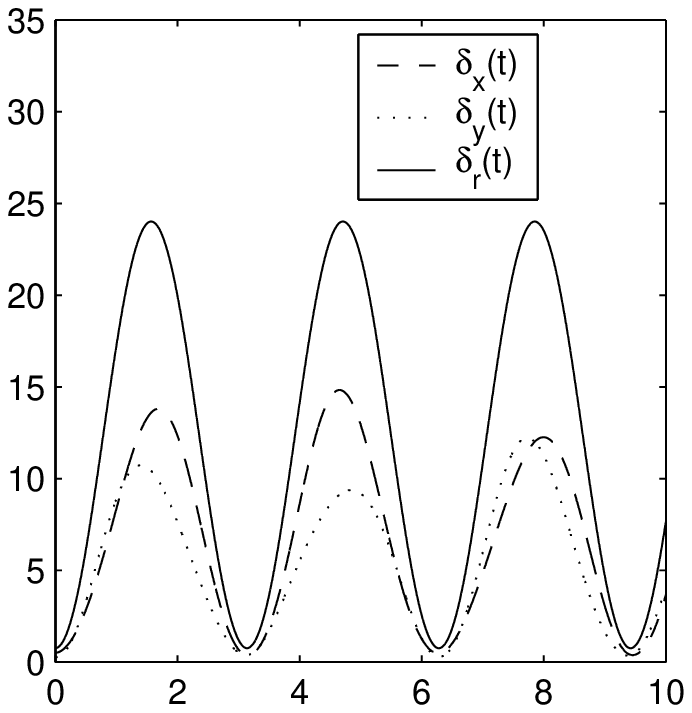,height=4cm,width=6cm,angle=0}}
\bigskip
\centerline{
(c)\psfig{figure=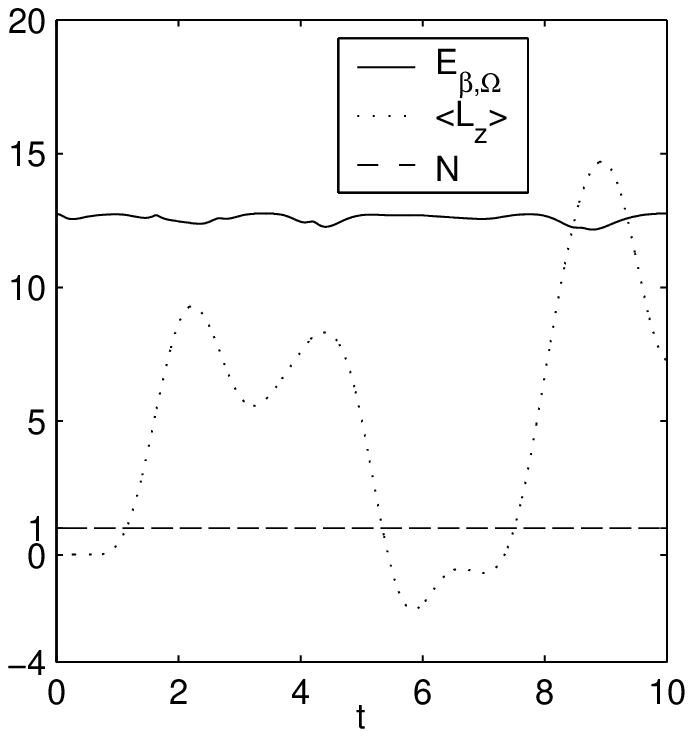,height=4cm,width=6cm,angle=0}
\qquad (d) \psfig{figure=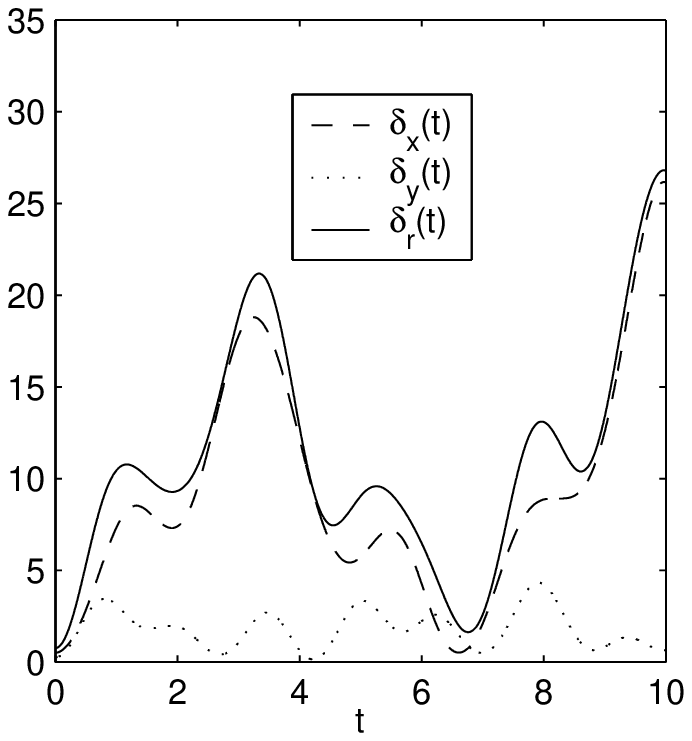,height=4cm,width=6cm,angle=0}}

\centerline{(e)\psfig{figure=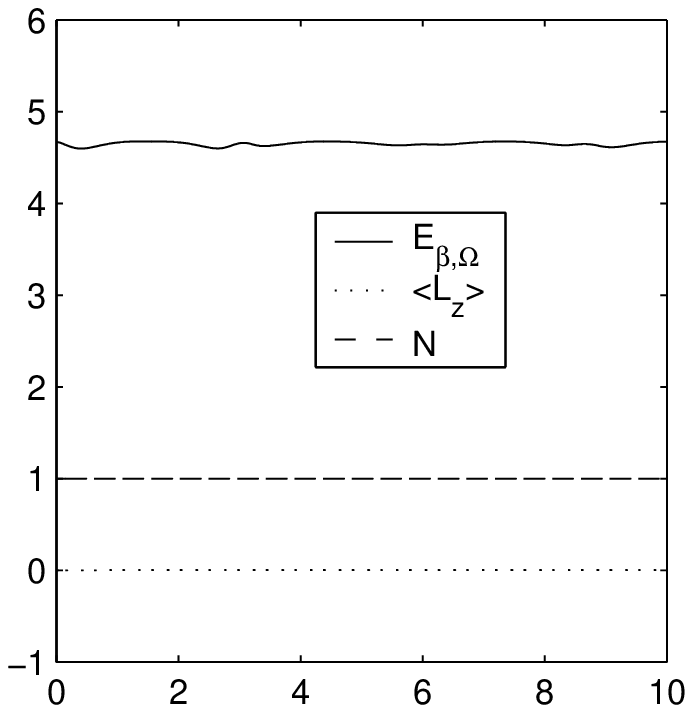,height=4cm,width=6cm,angle=0}
\qquad (f)
\psfig{figure=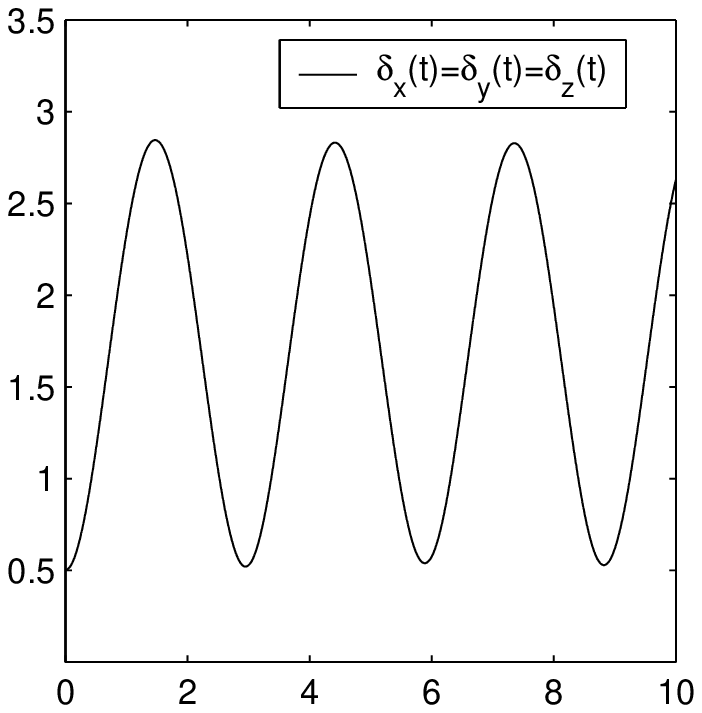,height=4cm,width=6cm,angle=0}}
\bigskip
\centerline{
(g)\psfig{figure=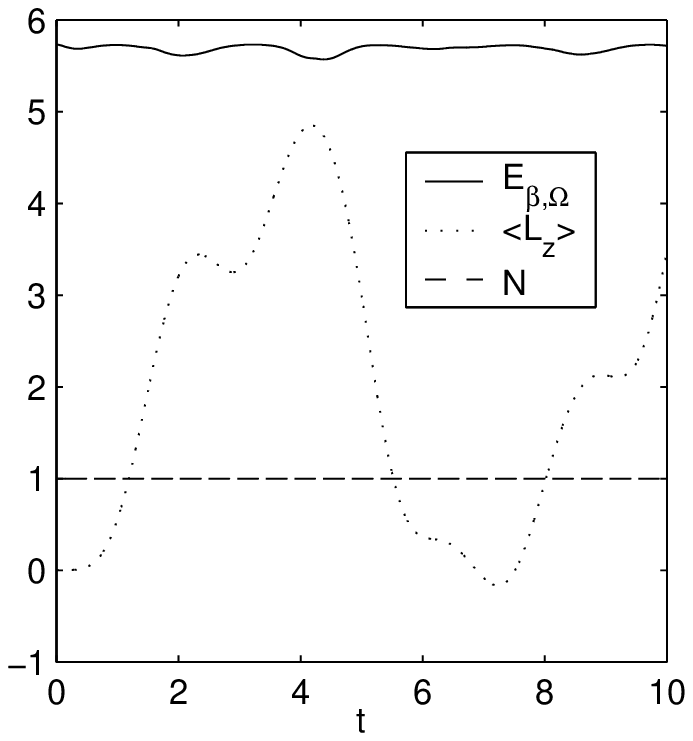,height=4cm,width=6cm,angle=0}
\qquad (h)
\psfig{figure=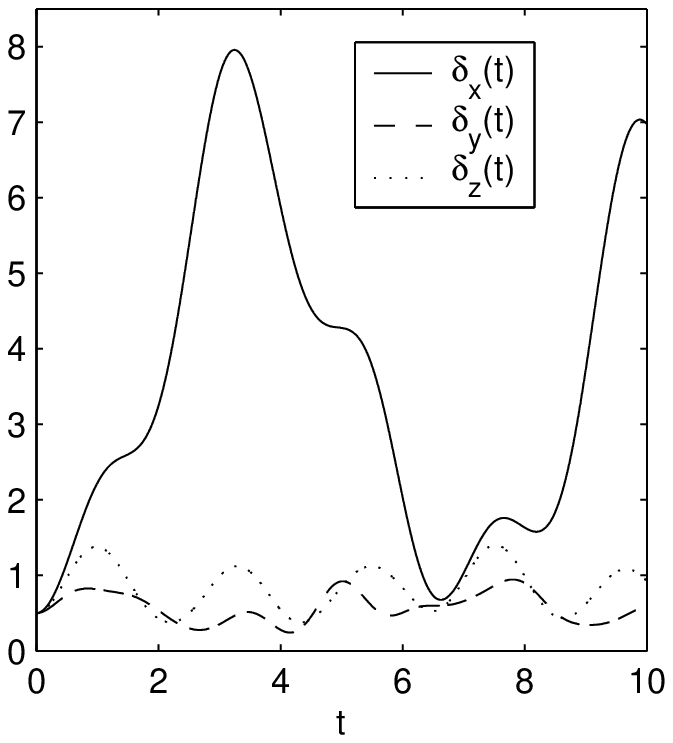,height=4cm,width=6cm,angle=0}}

Fig. 1: Time evolution of the normalization $N(t) :=
N(\psi)(t)$, energy $E_{\bt,\Og}(\psi)(t)$, angular momentum
expectation $\langle L_z\rangle(t)$ (left column), and condensate
widths $\dt_x(t)$, $\dt_y(t)$ and $\dt_r(t)$ (right column).
Results in 2D for $\gm_x=\gm_y=1$ (a\&b), and
$\gm_x=1$ and $\gm_y=2$ (c\&d); and
results in 3D for $\gm_x=\gm_y=\gm_z=1$ (e\&f), and
$\gm_x=1$, $\gm_y=2$ and $\gm_z=1.5$ (g\&h).
\end{figure}

From Fig. 1, we can see that: (i). The normalization is conserved
in both cases which confirms (\ref{mass}). (ii). The energy is not
conserved in the discretized level,
but the perturbation is very small, e.g. less than
$5\%$ (c.f. (a), (c), (e) and (g) of Fig. 1). (iii). The angular momentum
expectation is conserved when $\gm_x=\gm_y=1$ (c.f. (a) and (e) of Fig. 1)
which confirms the
analytical result (\ref{consang}), and oscillates when $1=\gm_x\ne
\gm_y=2$ (c.f. (c) and (g) of Fig. 1). (iv) The condensate widths
$\dt_x(t)$, $\dt_y(t)$ and $\dt_r(t)$ are periodic functions when
$\gm_x=\gm_y=1$ which confirm the analytical result
(\ref{solutiondtr}) (c.f. (b)  of Fig. 1), and are not
periodic functions when $1=\gm_x\ne \gm_y=2$ (c.f. (d) and (h) of Fig. 1).

 \subsection{Dynamics of a vortex lattice in rotating BEC}

In this subsection we numerically study the dynamics of a
 vortex lattice in rotating BEC by changing trap frequencies.
This study was motivated by the recent experiment \cite{Engels}
in which the frequencies of trapping potential of a stable BEC
were changed \cite{Engels}.
One of the most striking observation in the experiment
is that the condensate contains sheet-like
structures rather than individual vortex cores
in the dynamics by deforming the static trap \cite{Engels}.
By using the hydrodynamic forms of the GPE (\ref{gpeg})
in Thomas-Fermi regime, Cozzin et al. \cite{Cozzini} tried to
study this phenomena theoretically. But they  did not find
the sheet-like structures by changing the trap frequencies
in their theoretical study. Here we study this phenomena
by directly simulating the GPE (\ref{gpeg})
using our new numerical method.

We take $d=2$, $\bt_2=100$
and $\Og=0.99$ in (\ref{gpeg}).  The initial data $\psi_0(\bx)$ in
(\ref{gpegp}) is chosen as the ground state of (\ref{gpeg}) with
$d=2$, $\Og=0.99$, $\beta_2=100$ and $\gm_x=\gm_y=1$, which is
computed numerically by the normalized gradient flow proposed in \cite{Bao5}.
In the ground state, there are
about $61$ vortices in the vortex lattice (c.f. Fig. 2).
We solve the problem on
$\Og_\bx=[-24,24]\tm[-24,24]$ with mesh size $\btu x=\btu y= 3/16$
and time step $\btu t=0.001$.

  First we study free expansion of the quantized vortex lattice.
  In general, the size of a stable
vortex lattice in a BEC is too small to visualize it.
In experiments, by removing the
trap, i. e., letting the vortex lattice freely expands,
one can obtain an enlarged vortex lattice so as to take a photo for it
\cite{Engels}. Of course, they hope the vortex structure doesn't
change during the free expansion. Thus theoretical study of
free expansion is very helpful for experiments.
We start with the stable BEC and remove
the trapping at time $t=0$, i.e. choosing $\gm_x=\gm_y=0$ in
(\ref{gpeg}). Similar numerical study was
also carried out in \cite{Adhikari1} by a different numerical
method with much less number of vortices in the lattice.
Fig. 2 shows image plots of the
density $|\psi(\bx,t)|^2$ at different times
for the free expansion of the vortex lattice.
From the figure, we can see that when the trap is removed at $t=0$, the
vortex lattice will expand  with time and the  vortex structure as
well as  the rotational symmetry is kept during the expansion.
This gives a numerical justification for the free expansion used
in BEC experiments.

 \begin{figure}[htb] \label{fig3}
\centerline{\psfig{figure=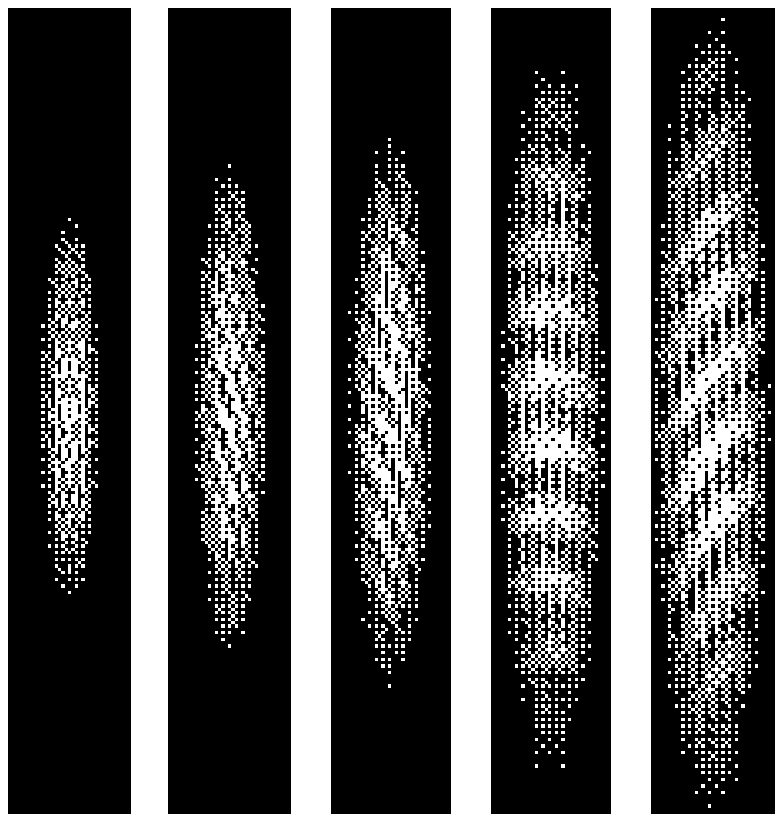,height=2cm,width=12cm,angle=0}}

Fig. 2: Image plots of the density
$|\psi(\bx,t)|^2$ on $[-18,18]\tm[-18,18]$
at different times $t=0$, $0.75$, $1.5$, $2.0$
and $2.75$ (from left to right) for the free expansion of a
quantized vortex lattice.
\end{figure}

Next, we study dynamics of the quantized vortex lattice by
changing the trap frequencies.  We study six different cases:
I. $\gm_x=1$, $\gm_y=1.5$; II. $\gm_x=1$, $\gm_y=0.75$;
III. $\gm_x=1.5$, $\gm_y=1$; IV. $\gm_x=0.75$, $\gm_y=1$;
V. $\gm_x=\sqrt{1.2}$,
$\gm_y=\sqrt{0.8}$; VI. $\gm_x=\sqrt{1.4}$,
$\gm_y=\sqrt{0.6}$. Similar numerical study was
also carried out in \cite{Adhikari1} by a different
method with much less number of vortices in the lattice.
Fig. 3  shows image plots of the
density $|\psi(\bx,t)|^2$ at
different times for cases I-II for changing frequencies in $y$-direction
only. Fig. 4 shows similar results
for cases III-IV for changing frequencies in $x$-direction
only, and Fig. 5  for cases V-VI
for changing frequencies in both $x$- and $y$-directions.

\begin{figure}[htb] \label{fig4}
\centerline{(a)\psfig{figure=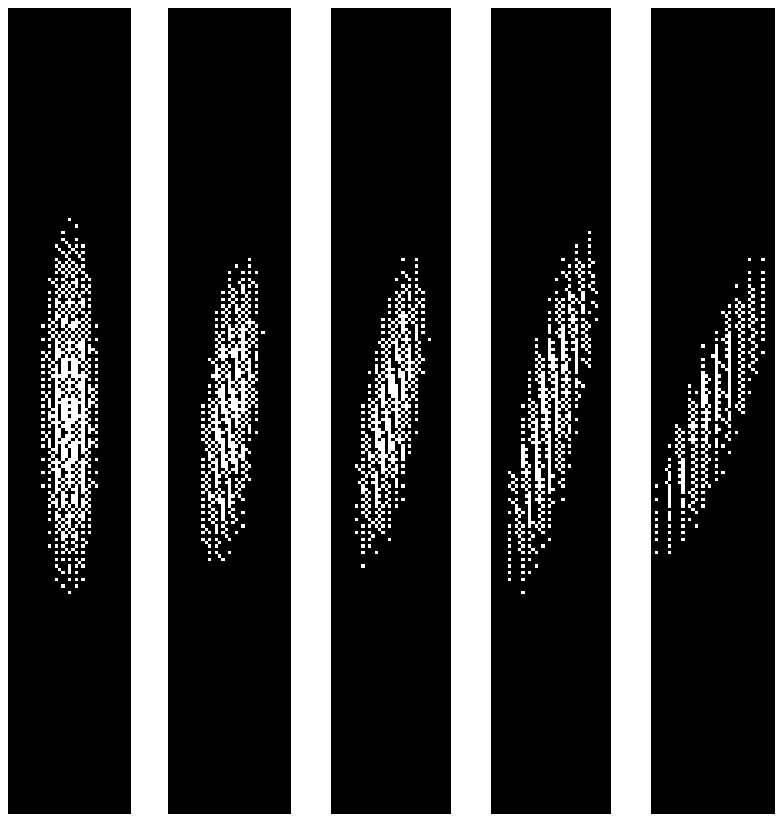,height=2cm,width=12cm,angle=0}}
\bigskip
\centerline{(b)\psfig{figure=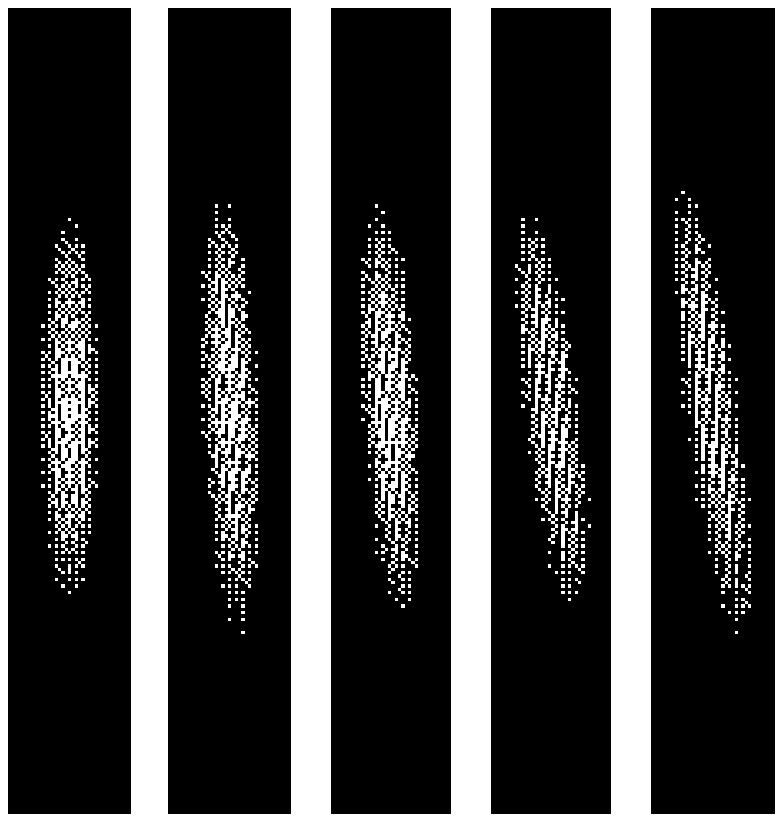,height=2cm,width=12cm,angle=0}}

Fig. 3: Image plots of the density $|\psi(\bx,t)|^2$ on
$[-18,18]\tm[-18,18]$  at different times t=$0$, $2.0$,
$4.0$, $6.0$, and $8.0$ (from left to right) by changing
frequency in $y$-direction only from $\gm_y=1$ to: (a)
$\gm_y=1.5$; (b) $\gm_y=0.75$.

\end{figure}

\begin{figure}[htb] \label{fig44}
\centerline{(a)\psfig{figure=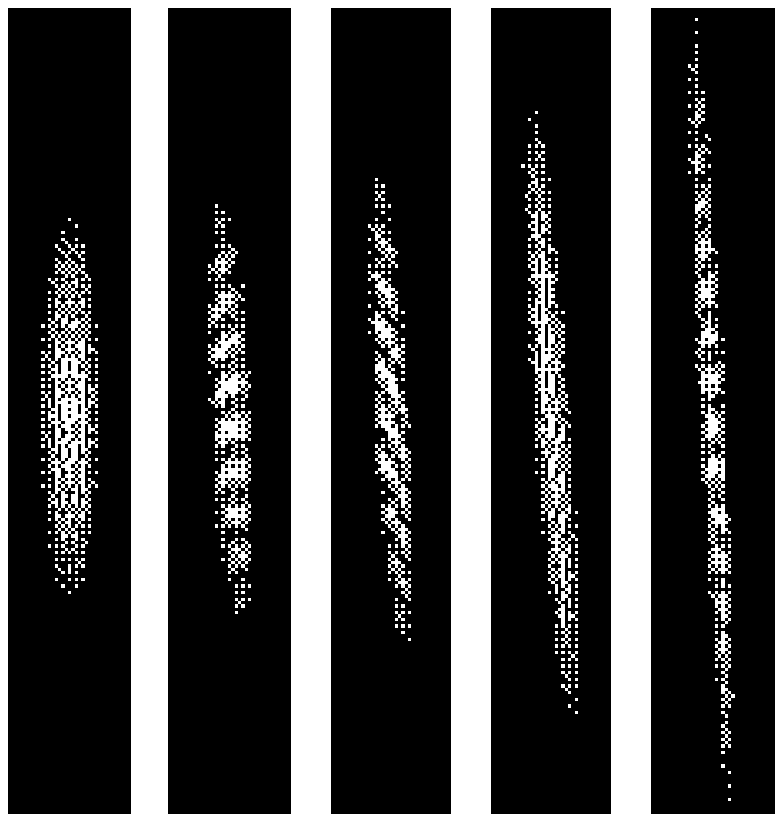,height=2cm,width=12cm,angle=0}}
\bigskip
\centerline{(b)\psfig{figure=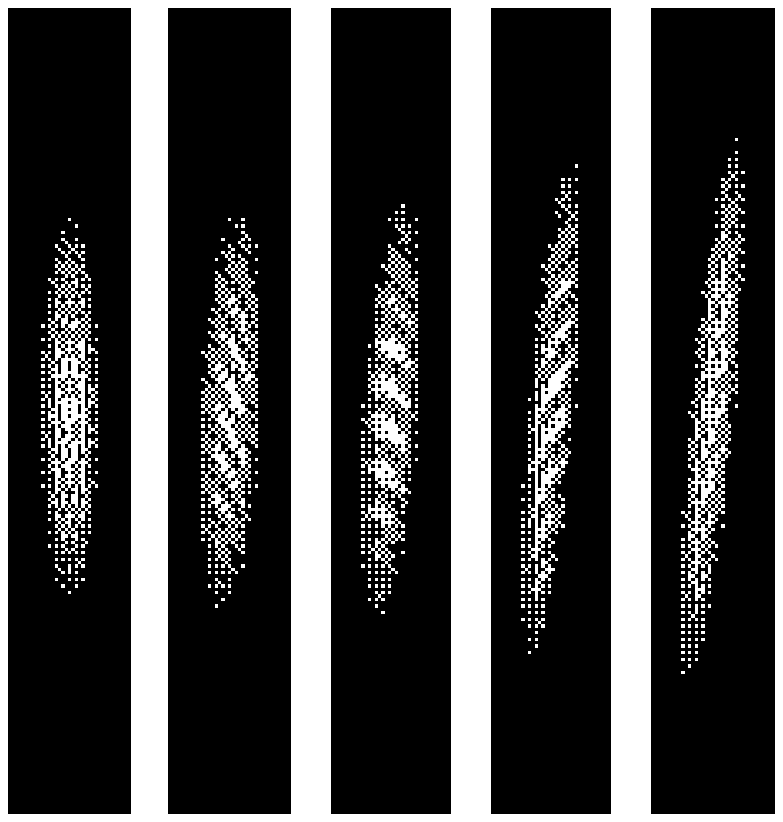,height=2cm,width=12cm,angle=0}}

Fig. 4: Image plots of the density $|\psi(\bx,t)|^2$ on
$[-18,18]\tm[-18,18]$  at different times t=$0$, $2.0$,
$4.0$, $6.0$, and $8.0$ (from left to right) by changing
frequency in $x$-direction only from $\gm_x=1$ to: (a)
$\gm_x=1.5$; (b) $\gm_x=0.75$.

\end{figure}

\begin{figure}[t!]  \label{fig5}
\centerline{(a)\psfig{figure=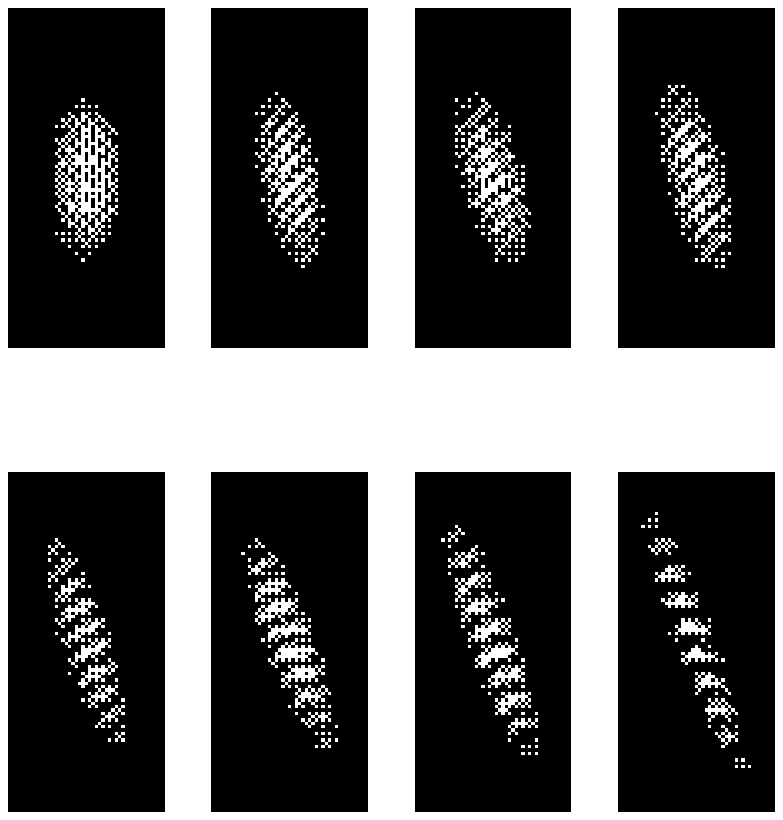,height=5cm,width=10cm,angle=0}}
\bigskip
\centerline{(b)\psfig{figure=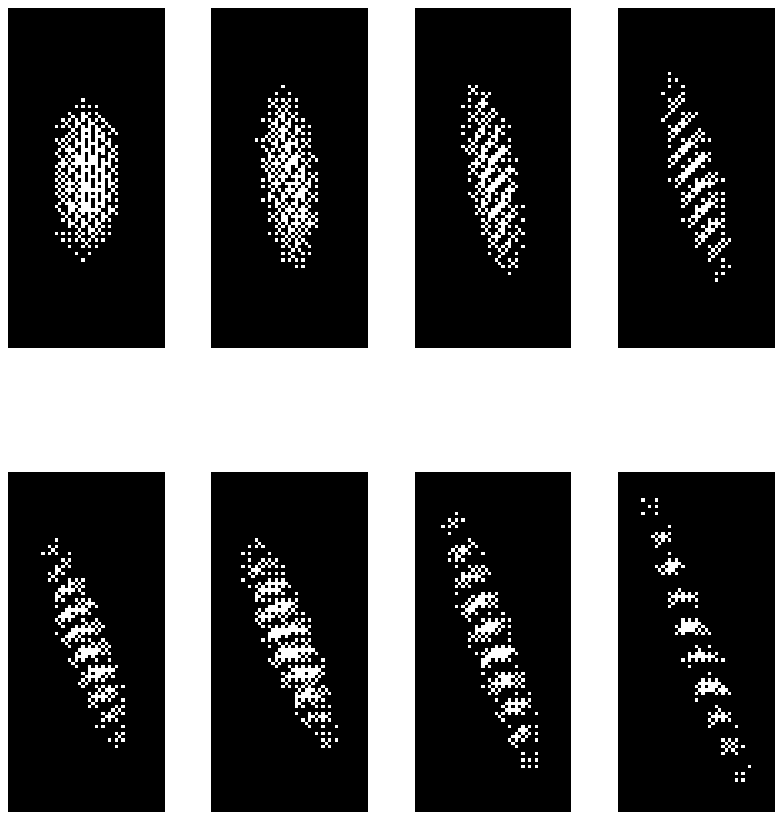,height=5cm,width=10cm,angle=0}}

Fig. 5: Image plots of the density $|\psi(\bx,t)|^2$ on
$[-18,18]\tm[-18,18]$  at different times by changing
trapping frequencies from $\gm_x=1$, $\gm_y=1$
to $\gm_x=\sqrt{1+\epsilon}$, $\gm_y=\sqrt{1.0-\epsilon}$.
(a) With $\epsilon=0.2$ for times
 $t=0$, $2$, $3$, $4$, $5$, $6$, $7$ and $8$ (from left to right);
(b) with $\epsilon=0.4$ for times $t=0$, $1.0$, $1.5$, $2.0$,
$2.5$, $3.0$, $4.0$ and $4.5$.

\end{figure}

   In Figs. 3-5, initially the condensate is assumed to be in
its ground state which is a vortex lattice with about
$61$ vortices. From the numerical results presented here,
 when the trap frequencies are changed at $t=0$, we find that: (i) Cases
I \& II correspond to changing trap frequency in
$y$-direction only. The condensate initially starts to contract
(c.f. (a) of Fig.
3) or expand (c.f. (b) of Fig. 3) in $y$-direction
 since the trap  frequency in $y$-direction is
increasing or decreasing at $t=0$. (ii) Cases III \& IV correspond to
changing trap frequency in $x$-direction only. Similar results
are observed (c.f. (a) and (b) of Fig. 4). (iii)
 Cases V \& VI correspond to
increasing and decreasing the  trapping frequencies in $x$ and
$y$-directions by the same value, i.e., $\epsilon$,
 respectively \cite{Cozzini}. The
condensate initially starts to contract and expand in  $x$ and
$y$-directions respectively (c.f. Fig. 5).
(iv) We numerically observed  the remarkable sheet-like vortices
in our numerical results (c.f. Figs. 3\&5).
One can compare our numerical results in Figs. 3\&5 with the
experimental results, e.g. Fig. 4 in \cite{Engels},
 and find very good qualitative
agreement in sheet-like vortex lattice formation.
Furthermore, we also found that when $\epsilon$ is bigger
in Cases V\&VI, the sheet-like vortices appear  earlier.

\subsection{Generation of giant vortex in rotating BEC}

In this subsection we numerically generate a giant vortex
in rotating BEC from its ground state by introducing
a localized loss term \cite{Engels1}. This study was motivated by
the recent experiment \cite{Engels1} and theoretical study
\cite{Simula} in which the giant vortex formation arises as a dynamics
effect. We take $d=2$, $\bt_2=100$, $V_p(\bx)\equiv0$
and $\Og=0.99$ in (\ref{GPE2}).  The initial data $\psi_0(\bx)$ in
(\ref{initial_data2}) is chosen as the ground state of (\ref{gpeg}) with
$d=2$, $\Og=0.99$, $\beta_2=100$ and $\gm_x=\gm_y=1$, which is
computed numerically by the normalized gradient flow proposed in \cite{Bao5}.
The localized loss term in (\ref{GPE2}) is chosen as a Gaussian
function of the form \cite{Simula}
\be
\label{loss}
W(x,y)=w_0 \exp\left[-\frac{(x-x_0)^2+(y-y_0)^2}{r_0^2}\right],
\qquad (x,y)\in {\Bbb R}^2,
\ee
where $w_0$, $x_0$, $y_0$ and $r_0$ are constants.
We take $w_0=1$ and $r_0=\sqrt{7}/2$ in (\ref{loss}) and
solve the problem (\ref{GPE2})-(\ref{initial_data2}) on
$\Og_\bx=[-24,24]\tm[-24,24]$ with mesh size $\btu x=\btu y= 3/16$
and time step $\btu t=0.001$.
Fig. 6 shows image plots of the density $|\psi(\bx,t)|^2$ at different
times for different values of $(x_0,y_0)$.

\begin{figure}[t!]  \label{fig7}
\centerline{(a)\psfig{figure=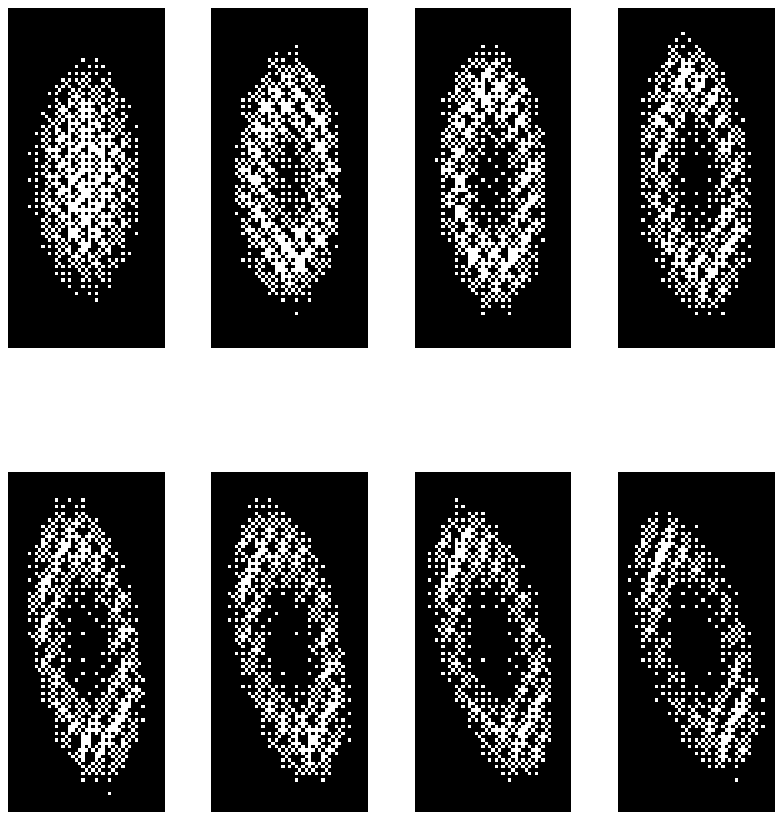,height=5cm,width=10cm,angle=0}}
\bigskip
\centerline{(b)\psfig{figure=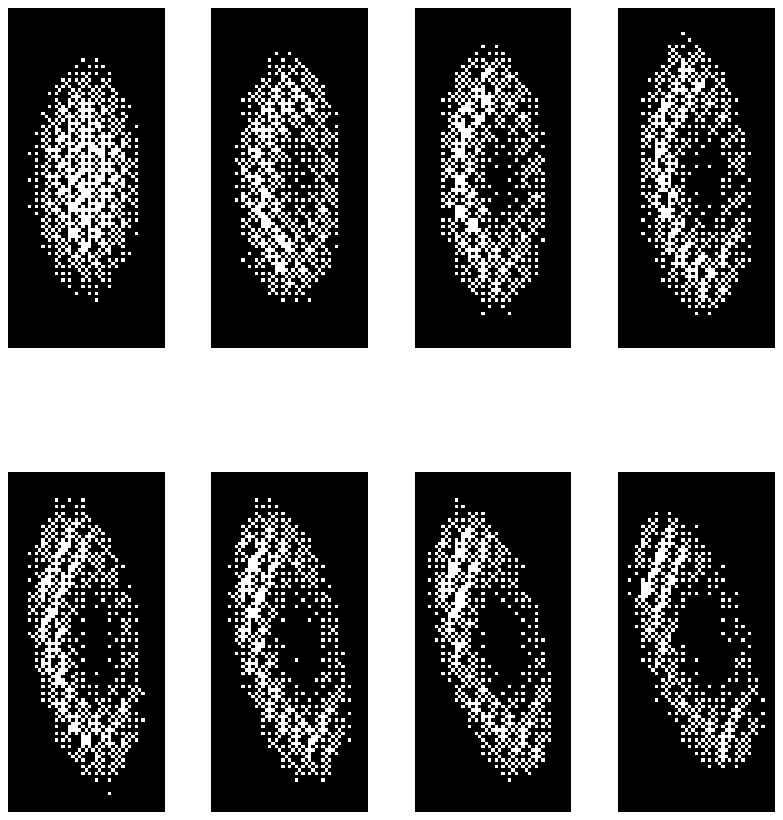,height=5cm,width=10cm,angle=0}}
\bigskip
\centerline{(c)\psfig{figure=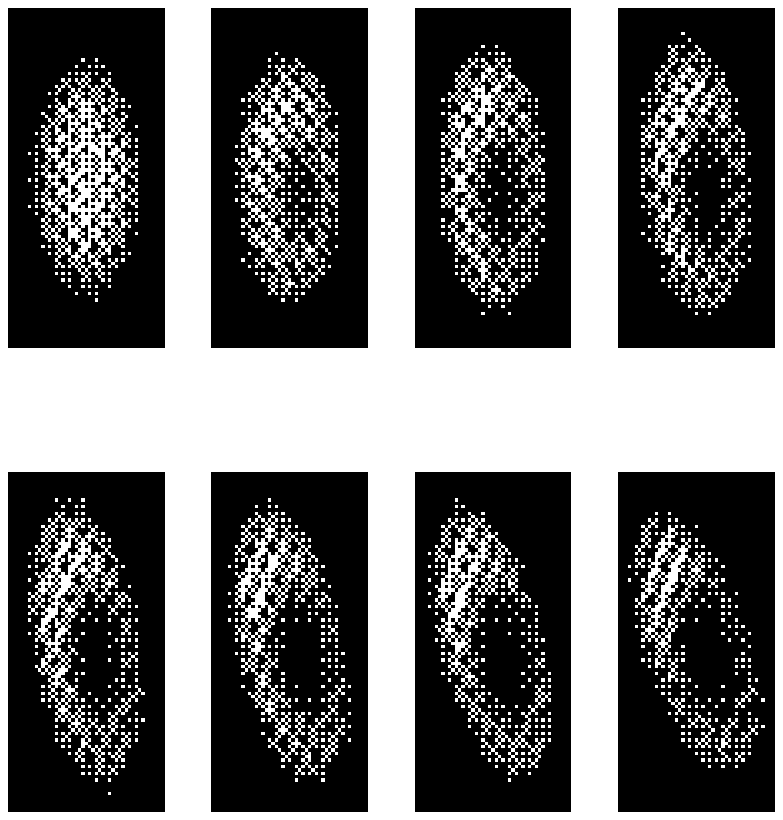,height=5cm,width=10cm,angle=0}}

Fig. 6: Image plots of the
density $|\psi(\bx,t)|^2$ on $[-12,12]\tm[-12,12]$  at different times
 t=$0$, $0.5$, $0.75$, $1.0$, $1.25$, $1.5$, $1.75$, and $2.0$
(from left to right) with $w_0=1$ and $r_0=\frac{\sqrt{7}}{2}$
in (\ref{loss})
for generating giant vortices. (a) $x_0=0$, $y_0=0$;
(b) $x_0=1.5$, $y_0=0$; (c) $x_0=1.5$, $y_0=1$.
\end{figure}

From Fig. 6, we can see that the giant vortex lattice is generated
due to the dynamic effect in a rotating BEC. The center of the
giant vortex is the same as the center of the localized loss term
and the size of the giant vortex depends on the values of $r_0$
and $w_0$. One can compare our numerical results in Fig. 6 with
the experimental results, e.g. Fig. 1 in \cite{Engels1}, and the
theoretical study, e.g. Fig. 1 in \cite{Simula}, and find very
good qualitative agreement in giant-vortex formation.

\section{Conclusions}

   We have proposed a new  time-spitting Fourier pseudospectral
method for computing dynamics of rotating BEC based on an
efficient approximation of GPE with an angular momentum rotation
term. The new method is explicit, unconditionally stable, and of
spectral accuracy in space and second order accuracy in time. It is time
reversible and  time transverse invariant in the discretized
level, just as the original GPE does, and
conserves the total density in the discretized level.
The efficient and accurate
numerical method was applied to study dynamics of a quantized
vortex lattice in rotating BEC. In the future, we plan to extend
the idea for constructing the new numerical method for one
component rotating BEC to multi-component rotating BEC and spinor
dynamics in a rotational frame, and apply the method to study
vortex line dynamics in rotating BEC in 3D.

\bigskip
\bigskip

\renewcommand{\theequation}{\Alph{section}.\arabic{equation}}
\begin{center}
{\bf Appendix A.  Discretization in 2D }
\end{center}
\setcounter{section}{1}
\setcounter{equation}{0}

For each fixed $y$, the operator in the equation (\ref{fstep1}) is
in  $x$-direction with constant coefficients and thus we can
discretize it in $x$-direction by a Fourier pseudospectral method.
Assume \be\label{expanx} \psi(x,y,t) = \sum_{p=-M/2}^{M/2-1}
\widehat{\psi}_p(y,t) \; \exp[i\mu_p (x-a)], \ee
 where
$\mu_p=\frac{2p\pi}{b-a}$ and $\widehat{\psi}_p(y,t)$ is the
Fourier coefficient for the $p$-th mode in $x$-direction. Plugging
(\ref{expanx}) into (\ref{fstep1}), noticing the orthogonality of
the Fourier functions, we obtain for $-\frac{M}{2}\le p \le
\frac{M}{2}-1$ and $c\le y\le d$: \be\label{odex}
i\;\p_t\widehat{\psi}_p(y,t) = \left(\frac{1}{2} \mu_p^2 +\Og y
\mu_p\right)\widehat{\psi}_p(y,t), \qquad t_n\le t \le t_{n+1}.
\ee The above linear ODE can be integrated in time {\sl exactly}
and we obtain \be \label{odexs}
\widehat{\psi}_p(y,t)=\exp\left[-i\left(\frac{1}{2} \mu_p^2 +\Og y
\mu_p\right)(t-t_n)\right]\ \widehat{\psi}_p(y,t_n), \qquad t_n\le
t \le t_{n+1}. \ee Similarly, for each fixed $x$, the operator in
the equation (\ref{fstep2}) is in  $y$-direction with constant
coefficients and thus we can discretize it in $y$-direction by a
Fourier pseudospectral method. Assume \be\label{expany}
\psi(x,y,t) = \sum_{q=-N/2}^{N/2-1} \widehat{\psi}_q(x,t) \
\exp[i\ld_q (y-c)], \ee where $\ld_q=\frac{2q\pi}{d-c}$ and
$\widehat{\psi}_q(x,t)$ is the Fourier coefficient for the $q$-th
mode in $y$-direction. Plugging (\ref{expany}) into
(\ref{fstep2}), noticing the orthogonality of the Fourier
functions, we obtain for $-\frac{N}{2}\le q \le \frac{N}{2}-1$ and
$a\le x\le b$: \be\label{odey} i\;\p_t\widehat{\psi}_q(x,t) =
\left(\frac{1}{2} \ld_q^2 -\Og x
\ld_q\right)\widehat{\psi}_q(x,t), \qquad t_n\le t \le t_{n+1}.
\ee Again the above linear ODE can be integrated in time {\sl
exactly} and we obtain \be \label{odeys}
\widehat{\psi}_q(x,t)=\exp\left[-i\left(\frac{1}{2} \ld_q^2 -\Og x
\ld_q\right)(t-t_n)\right]\ \widehat{\psi}_q(x,t_n), \qquad t_n\le
t \le t_{n+1}.
 \ee
From time $t=t_n$ to $t=t_{n+1}$, we combine the splitting steps
via the standard second order Strang splitting
\cite{Strang,Bao8,Bao9}: \bea \label{tssp2d}
\psi_{jk}^{(1)}&=&\sum_{p=-M/2}^{M/2-1}
 e^{-i\btu t(\mu_p^2
+2\Og y_k \mu_p)/4}\; \widehat{(\psi_k^n)}_p\; e^{i\mu_p(x_j-a)},
    \ 0\le j \le M, \quad 0\le k \le N, \nn\\
\psi_{jk}^{(2)}&=&\sum_{q=-N/2}^{N/2-1}
 e^{-i\btu t(\ld_q^2
-2\Og x_j \ld_q)/4}\; \widehat{(\psi_j^{(1)})}_q\;
e^{i\ld_q(y_k-c)},
    \ 0\le k \le  N, \quad 0\le j \le M, \nn\\
\psi^{(3)}_{jk}&=&\left\{\ba{ll}
e^{-i\btu t [V(x_j,y_k)+\bt_2|\psi_{jk}^{(2)}|^2]}\;\psi_{jk}^{(2)},
  &W(x_j,y_k)=0, \\
\frac{\psi_{jk}^{(2)}}{e^{\btu t W(x_j,y_k)}}
e^{-i[\btu t V(x_j,y_k)+\beta_2|\psi_{jk}^{(2)}|^2
(1-e^{-2\btu t W(x_j,y_k)})/2W(x_j,y_k)]}, &W(x_j,y_k)>0,\\
\ea\right. \nn \\
\psi_{jk}^{(4)}&=&\sum_{q=-N/2}^{N/2-1}
 e^{-i\btu t(\ld_q^2
-2\Og x_j \ld_q)/4}\; \widehat{(\psi_j^{(3)})}_q\;
e^{i\ld_q(y_k-c)},
    \ 0\le k \le  N, \quad 0\le j\le  M, \nn\\
\psi_{jk}^{n+1}&=&\sum_{p=-M/2}^{M/2-1}
 e^{-i\btu t(\mu_p^2
+2\Og y_k \mu_p)/4}\; \widehat{(\psi_k^{(4)})}_p\;
e^{i\mu_p(x_j-a)},
    \ 0\le j \le  M, \ 0\le k\le  N;\qquad \quad
\eea where for each fixed $k$, $\widehat{(\psi_k^\alpha)}_p$
($p=-M/2,\cdots,M/2-1$) with $\ap$ an index,
the Fourier coefficients of the vector
$\psi_k^\ap=(\psi_{0k}^\ap$, $\psi_{1k}^\ap$, $\cdots$,
$\psi_{(M-1)k}^\ap)^T$, are defined as \be \label{Fouv1}
\widehat{(\psi_k^\ap)}_p=\frac{1}{M}\sum_{j=0}^{M-1}
 \psi^\ap_{jk}\;e^{-i\mu_p(x_j-a)},  \quad
 p=-\fl{M}{2},\cdots,\fl{M}{2}-1;
\ee similarly, for each fixed $j$, $\widehat{(\psi_j^\ap)}_q$
($q=-N/1,\cdots,N/2-1$), the Fourier coefficients of the vector
$\psi_j^\ap=(\psi_{j0}^\ap$, $\psi_{j1}^\ap$, $\cdots$,
$\psi_{j(N-1)}^\ap)^T$, are defined as \be \label{Fouv2}
\widehat{(\psi_j^\ap)}_q=\frac{1}{N}\sum_{k=0}^{N-1}
 \psi^\ap_{jk}\;e^{-i\ld_q(y_k-c)},  \quad
 q=-\fl{N}{2},\cdots,\fl{N}{2}-1.
\ee

For the algorithm (\ref{tssp2d}) presented in Appendix A,
the total memory requirement is $O(MN)$ and the total
computational cost per time step is $O(MN\ln (MN))$. The scheme is
time reversible when $W(\bx)\equiv 0$,
just as it holds for the GPE (\ref{gpeg}), i.e. the
scheme is
unchanged if we interchange $n\leftrightarrow n+1$ and $\btu
t\leftrightarrow -\btu t$ in (\ref{tssp2d}).  Also, a main advantage of the
numerical method is its time-transverse invariance
when $W(\bx)\equiv 0$, just as it
holds for the GPE (\ref{gpeg}) itself. If a constant $\ap$ is
added to the external potential $V$, then the discrete wave
functions $\psi_{jk}^{n+1}$ obtained from (\ref{tssp2d}) get
multiplied by the phase factor $e^{-i\ap(n+1)\btu t}$, which
leaves the discrete quadratic observable  $|\psi_{jk}^{n+1}|^2$
unchanged. This property does not hold for the finite difference
scheme \cite{Kasamatsu,Wang}, the leap-frog spectral method
\cite{Zhang} and the efficient discretization proposed in
\cite{Bao3} for GPE with an angular momentum term.

\bigskip
\bigskip

 \begin{center}
{\bf Appendix B. Discretization in 3D}
\end{center}
\setcounter{section}{2}
\setcounter{equation}{0}

For each fixed $y$, the operator in the equation (\ref{fstep3}) is
in  $x$ and $z$-directions with constant coefficients and thus we
can discretize it in $x$ and $z$-directions by a Fourier
pseudospectral method. Similarly, for each fixed $x$, the operator
in the equation (\ref{fstep4}) is in  $y$ and $z$-directions with
constant coefficients and thus we can discretize it in $y$ and
$z$-directions by a Fourier pseudospectral method. The
discretizations of (\ref{fstep3}) and (\ref{fstep4}) are similar
as those for (\ref{fstep1}) and (\ref{fstep2}) respectively and
they are omitted here. For simplicity and convenience of the
reader, here we only present the algorithm for 3D GPE with an
angular momentum rotation term (\ref{gpeg}) with $0\le j\le M$,
$0\le k\le N$ and $0\le l\le L$: \bea \label{tssp3d}
\psi_{jkl}^{(1)}&=&\sum_{p=-M/2}^{M/2-1}\;\sum_{s=-L/2}^{L/2-1}
 e^{-i\btu t(2\mu_p^2+\gm_s^2
+4\Og y_k \mu_p)/8}\ \widehat{(\psi_k^n)}_{ps}\ e^{i\mu_p(x_j-a)}\
e^{i\gm_s(z_l-e)}, \nn\\
\psi_{jkl}^{(2)}&=&\sum_{q=-N/2}^{N/2-1}\;\sum_{s=-L/2}^{L/2-1}
 e^{-i\btu t(2\ld_q^2+\gm_s^2
-4\Og x_j \ld_q)/8}\ \widehat{(\psi_j^{(1)})}_{qs}\
e^{i\ld_q(y_k-c)}\ e^{i\gm_s(z_l-e)}, \nn\\
\psi^{(3)}_{jkl}&=&\left\{\ba{lr}e^{-i\btu t
[V(x_j,y_k,z_l)+\bt_3|\psi_{jkl}^{(2)}|^2]}\;\psi_{jkl}^{(2)},
 &W(x_j,y_k,z_l)=0,\\
\frac{\psi_{jkl}^{(2)}}{e^{\btu t W(x_j,y_k,z_l)}}
\exp\left[-i\left(\btu t V(x_j,y_k,z_l)+
\right.\right.
 &W(x_j,y_k,z_l)>0,\\
\left.\left.\beta_3|\psi_{jkl}^{(2)}|^2
(1-e^{-2\btu t W(x_j,y_k,z_l)})/2W(x_j,y_k,z_l)\right)\right],
\ea\right. \nn \\
\psi_{jkl}^{(4)}&=&\sum_{q=-N/2}^{N/2-1}\;\sum_{s=-L/2}^{L/2-1}
 e^{-i\btu t(2\ld_q^2+\gm_s^2
-4\Og x_j \ld_q)/8}\ \widehat{(\psi_j^{(3)})}_{qs}\
e^{i\ld_q(y_k-c)}\ e^{i\gm_s(z_l-e)}, \nn\\
\psi_{jkl}^{n+1}&=&\sum_{p=-M/2}^{M/2-1}\;\sum_{s=-L/2}^{L/2-1}
 e^{-i\btu t(2\mu_p^2+\gm_s^2
+4\Og y_k \mu_p)/8}\ \widehat{(\psi_k^{(4)})}_{ps}\
e^{i\mu_p(x_j-a)}\ e^{i\gm_s(z_l-e)}, \qquad \eea where for each
fixed $k$, $\widehat{(\psi_k^\ap)}_{ps}$ ($-M/2\le p\le M/2-1$,
$-L/2\le s\le L/2-1$) with $\ap$ an index,
the Fourier coefficients of the vector $\psi_{jkl}^\ap$
($0\le j< M$, $0\le l<L$), are defined as \[
\widehat{(\psi_k^\ap)}_{ps}=\frac{1}{ML}\sum_{j=0}^{M-1}
\sum_{l=0}^{L-1}
 \psi^\ap_{jkl}\ e^{-i\mu_p(x_j-a)} \ e^{-i\gm_s(z_l-e)},  \
 -\fl{M}{2}\le p<\fl{M}{2}, \ -\frac{L}{2}\le s<\frac{L}{2};
\] similarly, for each fixed $j$, $\widehat{(\psi_j^\ap)}_{qs}$
($-N/1\le q \le N/2-1$, $-L/2\le s\le L/2-1$) with $\ap$ an index,
the Fourier
coefficients of the vector $\psi_{jkl}^\ap$ ($k=0,\cdots,N$,
$l=0,\cdots,L$), are defined as \[
\widehat{(\psi_j^\ap)}_{qs}=\frac{1}{NL}\sum_{m=0}^{N-1}
\sum_{l=0}^{L-1}
 \psi^\ap_{jkl}\ e^{-i\ld_q(y_k-c)} \ e^{-i\gm_s(z_l-e)},  \
 -\fl{N}{2}\le q<\fl{N}{2}, \ -\frac{L}{2}\le s < \frac{L}{2};
 \] with $\gm_s=\frac{2\pi s}{f-e}$ for
$s=-L/2,\cdots, L/2-1$.

For the discretization in 3D, the total memory requirement is
$O(MNL)$ and the total computational cost per time step is
$O(MNL\ln (MNL))$. Furthermore,
 the discretization is time
reversible and time transverse invariant in the discretized
level when $W(\bx)\equiv 0$.

\bigskip

\begin{center}
{\large \bf Acknowledgment}
\end{center}
The authors acknowledge support  by the National University of Singapore
 grant No. R-151-000-035-112 and the referees for their valuable
comments and suggestions to improve the paper.
W.B. also thanks hospitality during
his extended visit at Department of Mathematics, Capital Normal
University in Beijing where part of the work was carried out.

\end{document}